\documentclass{iopart}

\usepackage{amssymb}
\usepackage{amsthm}
\usepackage{graphicx}

\newtheorem{Th}{Theorem}
\newtheorem{lemma}{Lemma}
\newtheorem{cor}{Corollary}

\newtheorem{prop}{Proposition}

\begin{document}

\title{Supersymmetric transformations for coupled channels with
threshold differences}

\author{Boris F Samsonov$^1$, Jean-Marc Sparenberg$^2$ and Daniel Baye$^2$}
\address{$^1$ Physics Department, Tomsk State University,
36 Lenin Avenue, 634050 Tomsk, Russia}
\address{$^2$ Physique Quantique, C P 229,
Universit\'{e} Libre de Bruxelles, B 1050 Bruxelles, Belgium}
\eads{\mailto{samsonov@phys.tsu.ru}, \mailto{jmspar@ulb.ac.be} and
\mailto{dbaye@ulb.ac.be}}

\begin{abstract}
The asymptotic behaviour of the superpotential of
general SUSY transformations
for a coupled channel Hamiltonian with different thresholds
 is analyzed.
It is shown that asymptotically the superpotential can tend to a
diagonal matrix with
an arbitrary number of
positive and negative entries
 depending on the choice of the factorization solution.
The transformation of the Jost matrix is
generalized to ``non-conservative" SUSY transformations introduced in
Sparenberg {\it et al} (2006 {\it J.\ Phys.\ A: Math.\ Gen.\ } {\bf 39} L639).
Applied to the zero initial potential the method
permits to construct superpartners
 with a nontrivially coupled Jost-matrix.
Illustrations are given for two- and three-channel cases.
\end{abstract}



\section{Introduction}

In the context of quantum scattering inverse problems,
not much is known about coupled-channel problems with threshold differences,
i.e., inelastic problems (see paragraph IX.4 of \cite{chadan:89}
and references therein).
Though well-known single-channel methods based on the
 Marchenko or Gel'fand-Levitan approaches
have been generalized to these coupled-channel problems,
no sufficiently simple  method for
inverting experimental data has been deduced from these
generalizations.
This may be explained by the fact
that in its general form the method requires
knowing the whole set of scattering data and, in particular,
the whole scattering (S) matrix should be known,
whereas only the open-channel submatrix is accessible
from experimental data.
Another reason is related with
the complicated character of the
Gel'fand-Levitan-Marchenko equation
which in this case was solved only numerically \cite{FG}.
In the one channel case the second difficulty was overcome with the help
of the supersymmetric quantum mechanics (SUSY QM) approach
(see e.g.\  \cite{cooper:01,valladolid:04})
which is equivalent to the Gel'fand-Levitan-Marchenko method when
the kernel of the integral equation is degenerate \cite{sukumar:85c,myJPA}.
Moreover, when applied to the zero initial potential,
these transformations lead to exactly-solvable
potentials with scattering matrices being arbitrary-order rational functions
of the wave number and hence providing excellent fits of experimental data
\cite{baye:04}.
Until recently, however,
generalizing such supersymmetric transformations of
the zero potential to inelastic coupled-channel problems
seemed to be impossible
 as a matter of principle
\cite{amado:88a,amado:88b,cannata:93}.
Thus, experimental data with coupling did not seem accessible to this method.

Fortunately, as it was recently announced \cite{sparenberg:06},
this strong limitation
is explained by an unnecessary condition imposed on
transformation functions used up to now.
Usual supersymmetric transformations
\cite{cooper:01,sukumar:85c,myJPA,baye:04,amado:88a,amado:88b,cannata:93}
can be described in terms of transformation
operators relating solutions of two Sturm-Liouville problems, and hence keeping
boundary conditions unchanged; this is why we call them ``conservative".
Changes in the spectrum due to such transformations may correspond only to
elements
from the kernel of either the transformation operator or its adjoint form.
This is just the reason why the spectrum of two supersymmetric
partners may differ
only by a finite number of levels.
Although the existence of transformations of another kind,
breaking boundary conditions,
is known for some time (see e.g.\ \cite{samsonov:98b}), their use in the
single-channel
case has been limited to a very specific application \cite{samsonov:03},
where usual
transformations happen to be sufficient.
In the inelastic coupled-channel case, on the contrary, these ``non-conservative"
transformations seem to be of fundamental importance
as they lead to exactly-solvable
potential models with non-trivially coupled $S$-matrices. We hope
that these new transformations may become a keystone for an
 inversion procedure both simple enough for practical realization
 and accurate enough to fit experimental data with a good
 precision.

This renewal of interest for supersymmetric transformations
of inelastic coupled-channel problems implies that a general
study of their properties is necessary.
In particular, the Jost-matrix transformation,
from which the transformed S-matrix can be deduced,
should be known in the general case.
In  \cite{sparenberg:06},
it is shown that the Jost-matrix modification depends on the value
of the superpotential matrix $U(r)$ at infinity, which in turn depends on the
asymptotic behaviour of the factorization solution.
In the single-channel case, $U(\infty)$ is a positive (resp.\ negative)
real number when the factorization solution increases (resp.\ decreases)
at infinity.
In the coupled-channel case, $U(r)$ is a matrix
and its asymptotic behaviour is much more complicated.
The main goal of the present work is to study carefully this
behaviour.

Section 2 reviews definitions from multichannel scattering theory.
As an introduction to
Section 3
we first review briefly
general properties of multichannel supersymmetric
transformations and then we analyze the maximal number of arbitrary
parameters entering into the superpotential and as a result into the
transformed potential. Next we
 discuss in a detailed way the superpotential asymptotic behaviour
 proving our main theorem and give a closed expression for the
 transformed Jost solution and Jost function.
In Section 4 we derive a general form of the superpotential which is
a SUSY partner of the zero potential.
Section 5 illustrates these findings by examples and comparisons
with existing results from the literature.
Section 6 contains conclusions and perspectives.


\section{Multichannel scattering}

Let us first define our notation and briefly recall some notions of scattering
theory \cite{taylor:72,newton:82} used below.
We consider a multichannel radial Schr\"odinger equation
that reads in reduced units
\begin{equation}\label{schr}
H\psi(k,r)=k^2\psi(k,r)\qquad
H=-\frac{\rmd^2}{\rmd r^2}+V
\end{equation}
where $r$ is the radial coordinate, $V$ is an $N\times N$ real symmetric matrix,
and $\psi$ may be either a matrix-valued or a vector-valued solution.
By $k$ we denote either a point in the space ${\mathbb C}^N$,
$k=\left\{k_1,\ldots,k_N\right\}$,
$k_i\in \mathbb C$ such that $\mbox{Im}\,k_i\geqslant0$,
or a diagonal matrix with the non-vanishing entries $k_i$,
$k=\mbox{diag}(k_1,\ldots,k_N)$.
The complex wave numbers $k_i$ are related to the center-of-mass energy $E$
and the channel thresholds $\Delta_1,\ldots, \Delta_N$,
which are supposed to be different from each other,
by
\begin{equation}
k_i^2=E-\Delta_i.
\end{equation}
We do not assume a fixed order of channels since
 any necessary order can be achieved
by a suitable
permutation of the rows in the Schr\"odinger equation
\eref{schr}.
For simplicity
we will limit ourselves to unequal thresholds here and defer the general
study of equal and unequal thresholds to a future work.
We will also assume potential $V$ to be short-ranged at infinity, i.e.,
there exists an $\varepsilon>0$ such that
\begin{equation}
\int_0^\infty \rme^{\varepsilon r}|V_{ij}(r)|dr<\infty
\end{equation}
where $V_{ij}$, $i,j=1,\ldots,N$ are entries of  matrix $V$.
Under such assumptions,
the Schr\"odinger equation has two $N\times N$ matrix-valued solutions $f(\pm k,r)$
(Jost solutions) such that
\begin{equation}\label{eAs}
f(\pm k,r) \mathop{\to}_{r\rightarrow\infty} \exp(\pm \rmi kr)=
\mbox{diag}[\exp(\pm \rmi k_1 r),\ldots,\exp(\pm \rmi k_N r)].
\end{equation}
The columns of these
matrices form a basis in the $2N$-dimensional  solution space
of the Schr\"odinger equation with a given value of $E$.
In general, these solutions are complex and satisfy the symmetry property
$f(k,r)=f^*(-k^*,r)$, where asterisk denotes complex conjugation;
for real energies below all thresholds, $k=-k^*$ and the Jost solutions are real.

Next we define the regular solution $\varphi(k,r)$
and the irregular solution $\eta(k,r)$
by their behaviour at the origin.
For the sake of simplicity, we limit ourselves to bounded $s$-wave potentials,
in which case these solutions satisfy
\begin{eqnarray}\label{regcond}
\varphi(k,0)=0 \qquad \varphi'(k,0)=I \\
\eta(k,0)=I \qquad \eta'(k,0)=0
\label{regcondE}
\end{eqnarray}
where prime means derivation with respect to $r$ and $I$
denotes the identity matrix.
This definition shows that the columns of these
matrices also form a basis in the  solution space
of the Schr\"odinger equation.
In terms of the Jost solutions, these solutions read
\begin{eqnarray}
\varphi(k,r) & = & \frac{1}{2\rmi}
\left[f(k,r)k^{-1}F(-k)-f(-k,r)k^{-1}F(k)\right]
\label{regsol} \\
\eta(k,r) & = & \frac{1}{2\rmi}
\left[f(k,r)k^{-1}G(-k)-f(-k,r)k^{-1}G(k)\right]
\label{irregsol}
\end{eqnarray}
where $F(k)$ is the Jost matrix
\begin{equation}\label{FJost}
F(k)=f^T(k,0)
\end{equation}
with $T$ meaning transposition, and matrix $G(k)$ is defined as
\begin{equation}\label{G}
G(k)=-[f'(k,0)]^T.
\end{equation}
Proving \eref{regsol} and \eref{irregsol} requires calculating,
both at the origin and at infinity,
the Wronskian
$W[\varphi(k,r),f(k,r)] \equiv \varphi^T(k,r) f'(k,r) - [\varphi'(k,r)]^T f(k,r)$,
which generalizes the usual definition of the Wronskian of one-component functions
to $N$ channels,
and the Wronskian $W[\eta(k,r),f(k,r)]$.
\Eref{schr} implies that the value of these Wronskians is independent of $r$,
as well as that of $W[f(-k,r),f(k,r)]=2\rmi k$.
For real energies,
both solutions $\varphi$ and $\eta$ are purely real because they satisfy
a system of differential equations with
real coefficients and
 real boundary conditions.
For energies below all thresholds or for energies above all thresholds,
this can also be directly checked on \eref{regsol} and \eref{irregsol},
using the symmetry properties $F(k)=F^*(-k^*)$ and $G(k)=G^*(-k^*)$.

The Jost matrix defines both scattering and bound states properties.
The scattering matrix, which is symmetric, reads
\begin{equation}\label{S}
S(k)=k^{-1/2}F(-k)F^{-1}(k)k^{1/2}=
k^{1/2}[F^{-1}(k)]^T F^T(-k)k^{-1/2}.
\end{equation}
Bound-state energies, $E=E_m$, $m=1,\ldots,M$,
correspond to zeros of the determinant of the Jost function,
$\det F(\rmi\kappa_m)\equiv 0$, such that
$\mbox{Re}\,\kappa_{m,i}>0$, $\mbox{Im}\,\kappa_{m,i}=0$ with
$E_m=-\kappa_{m,i}^2+\Delta_i$, $i=1,\ldots,N$
below all thresholds.
For potentials satisfying the above assumptions,
the number $M$ of bound states is finite.


\section{Multichannel SUSY transformations}

According to the multichannel SUSY approach \cite{amado:88a,amado:88b},
applying the transformation operator
\begin{equation}\label{Am}
A^-=-\frac{\rmd}{\rmd r}+U(r)
\end{equation}
to solutions $\psi(k,r)$ of \eref{schr} leads to solutions $\tilde\psi(k,r)$
of the new equation
\begin{equation}\label{tildeschr}
\tilde H\tilde\psi(k,r)=k^2\tilde\psi(k,r) \qquad
\tilde H=-\frac{\rmd^2}{\rmd r^2}+\tilde V(r)
\end{equation}
where $\tilde{V}$, like $V$, is supposed to be
a real, short-ranged, bounded and
 symmetric $N\times N$ matrix
and $\tilde\psi$, like $\psi$, may be either a matrix-valued
or a vector-valued function.
The matrix-valued function $U$ (usually called superpotential)
is expressed in terms of a matrix-valued solution of \eref{schr}
at a fixed value of $E=\cal{E}$ below all thresholds
(this parameter is known as the factorization constant),
which we denote $\sigma$ and call transformation function
or factorization solution.
Defining the corresponding wave number diagonal matrix $\kappa$
by its positive elements $\kappa_i=\sqrt{\Delta_i - \cal{E}}$, one has
\begin{equation}
H \sigma(r)=-\kappa^2 \sigma(r)
\end{equation}
and
\begin{equation}\label{U}
U(r)=\sigma'(r)\sigma^{-1}(r).
\end{equation}
Then for $E\ne\cal{E}$ one has
$
\tilde\psi=A^-\psi.
$
The specific form \eref{Am}, \eref{U} of the transformation operator $A^-$
results in the potential $\tilde V$ from \eref{tildeschr} being of the form
\begin{equation}\label{Vtilde}
\tilde V(r)=V(r)-2U'(r).
\end{equation}
To have a real and symmetric potential \eref{Vtilde} we restrict
$\sigma$ to be real and such that its self-Wronksian vanishes,
$W(\sigma,\sigma)=0$.

For $E=\cal{E}$ a particular solution of \eref{tildeschr} is
\begin{equation}
\tilde\phi(\kappa,r)=(\sigma^{T})^{-1}(r)\qquad
\tilde H \tilde\phi(\kappa,r)=-\kappa^2\tilde\phi(\kappa,r).
\end{equation}
Other solutions $\tilde\psi$ corresponding to the same $\cal{E}$ may be found as
usual from the property $W(\tilde\phi,\tilde\psi)=-I$, which gives
\begin{equation}
\fl \tilde\psi(\kappa,r)=(\sigma^{T})^{-1}(r)\int_{r_0}^r \sigma^T(s)\sigma(s)ds \qquad
\tilde H\tilde\psi(\kappa,r)=-\kappa^2\tilde\psi(\kappa,r).
\end{equation}

The conventional SUSY transformations have the property that if
$\psi(0)=0$ then
$\tilde\psi(0)=(A^-\psi)_{r=0}=0$
which requires some additional
limitation on the transformation function $\sigma(r)$.
Following \cite{sparenberg:06}
rejecting this limitation leads to loosing this property
of the transformation operator $A^-$.
Such transformation operators violate the vanishing behaviour of
the solution at the origin and we call them ``non-conservative".
Nevertheless, since $A^-$
transforms solutions of the initial differential equation into
solutions of the transformed equation the full information about
the new Hamiltonian $\tilde H$ is accessible.
In particular its
Jost function and $S$-matrix can be constructed explicitly.
Below
we will concentrate our attention mainly on non-conservative transformations
although our main result
(Theorem \ref{Th:Uinf} in Section \ref{Smain}) is valid
for the general case.

\subsection{Number of arbitrary parameters in
superpotential}

First we discuss the general form of the
transformation function $\sigma$ introduced in the previous section.
We notice that the general vector-valued solution of the
Schr\"odinger equation \eref{schr} contains $2N$ integration
constants and can always be presented as a linear combination
of $2N$ fixed linearly independent solutions.
In general,
transformation function $\sigma$ may be composed of $N$ such
solutions.
Therefore it may contain $2N^2$ arbitrary parameters at
most.
But
as far as the new potential \eref{Vtilde} is concerned
there is a big redundancy between these parameters.
Indeed,
because of the specific form of the superpotential \eref{U}
 a multiplication of the transformation
function $\sigma$
 on the right by a non-singular constant matrix
does not affect the superpotential.
A minimal set of arbitrary parameters is given by
\begin{Th}
\label{Th1}
Given the initial potential $V(r)$, fixed
thresholds and factorization energy,
the most general transformed potential $\tilde V(r)$
 is completely determined
by the value $U(0)$ of
the symmetric superpotential matrix at the origin.
It is calculated by formulas \eref{Vtilde} and \eref{U}
 where
\begin{equation}\label{sigU0}
\sigma(r)=\eta(\rmi\kappa,r)+\varphi(\rmi\kappa,r)U(0)
\end{equation}
and contains $N(N+1)/2$ arbitrary real parameters which are the entries
of matrix $U(0)$.
\end{Th}

\begin{proof}
The general matrix-valued solution of the Schr\"odinger equation
\begin{equation}\label{sigreg}
\sigma(r)=\eta(\rmi\kappa,r)C_1+\varphi(\rmi\kappa,r)D_1,
\end{equation}
with real matrices $C_1$ and $D_1$
produces the most general real superpotential \eref{U} and, hence,
potential \eref{Vtilde}.
Definitions \eref{regcond} and \eref{regcondE} then imply
that $\det\sigma(0)=\det C_1$
vanishes if and only if matrix $C_1$ is singular.
In this case $\sigma(0)$ is not invertible and the superpotential and
hence the transformed potential $\tilde V$ become singular at the origin;
as stated above, we want to avoid this case here and therefore
we impose the condition $\det C_1\ne0$.
It is now clear that in this case
matrix $C_1$ does not affect the superpotential $U$ as given in \eref{U}
since we can
multiply \eref{sigreg} by $C_1^{-1}$ on the right,
which leaves the superpotential unaffected
 or, equivalently without loosing generality, put $C_1=I$.
The superpotential $U$ and hence the transformed potential $\tilde V$
thus only depend on the $N^2$ parameters appearing in $D_1$.

This simplified writing allows us to express easily
the other condition imposed on
the real transformation function \eref{sigreg},
namely the symmetry of $U$ and $\tilde V$.
As mentioned above, this happens when $W(\sigma,\sigma)=0$,
which gives $N(N-1)/2$ equations for the elements of matrix $D_1$.
The value of this Wronskian being $r$-independent,
\eref{sigreg} can be used to calculate it at the origin with
 \eref{regcond} and \eref{regcondE}
which leads to
\begin{equation}
W[\sigma(r),\sigma(r)]=D_1-D_1^T=0.
\label{symor}
\end{equation}
The superpotential and transformed potential are thus symmetric when $D_1$
is chosen symmetric.
This can also be checked on the value of the superpotential at the origin
which reads, according to
\eref{regcond}, \eref{regcondE}, \eref{U} and \eref{sigreg},
$U(0)=D_1$.
\end{proof}

To calculate the Jost matrix for the transformed potential
according to \eref{FJost} we need to know its Jost solution which
is defined by the asymptotic behaviour \eref{eAs}. Usually a
supersymmetry transformation changes this behaviour.
So we have
to analyze the asymptotic behaviour of the function
$\tilde f(k,r)=A^-f(k,r)$ which is mainly defined by the asymptotics of
superpotential $U(r)$.


\subsection{Asymptotic behaviour of superpotential\label{Smain}}

According to \eref{U}, the asymptotic behaviour of the superpotential
depends on the asymptotic
behaviour of the factorization solution.
This time, in place of \eref{sigreg},
we choose to write the factorization solution as
\begin{equation}\label{sigJost}
\sigma(r)=f(-\rmi \kappa,r)\kappa^{-1/2}C_2+
f(\rmi\kappa,r)\kappa^{-1/2}D_2
\end{equation}
with $C_2$ and $D_2$ being some constant matrices;
factor $\kappa^{-1/2}$ is introduced for further convenience.
We notice that matrices $C_2$ and $D_2$ should satisfy the
 condition
 \begin{equation}\label{symDC}
D_2^T C_2 - C_2^T D_2 = 0
\end{equation}
 following from
the symmetry property of the superpotential
$W[\sigma(r),\sigma(r)]=0$.

We show below that the asymptotic behaviour of the
superpotential crucially depends on the structure of matrix
$C_2$.
In particular, it depends on the rank of $C_2$ and if
rank\,$C_2=R<N$ it depends on an interrelation between the values
of
 thresholds and linear dependence between rows of $C_2$.
This interrelation becomes more transparent for a specific order
of channels. Therefore before going further we will first
rearrange the channels
taking into account the structure of matrix
$C_2$.
Our main aim in this reordering is to collect together
both
 all linearly independent rows of matrix $C_2$ and
 its linearly independent columns.
As it was already noticed changing
channels corresponds to going to another starting Hamiltonian
\eref{schr}.
But evidently it corresponds to the same physical
system after the reordering.
The permutation of columns simultaneously both in $C_2$ and in $D_2$
 is equivalent to a
multiplication on the right
of the whole factorization solution $\sigma$ by a constant
non-singular matrix which evidently does not change the
superpotential $U$ as given by \eref{U}.

We rearrange the rows
of $C_2$ together with the corresponding channels in the
following way.
The first channel with wavenumber $\kappa'_1$
and, hence, the first row of the reordered matrix
(we denote $C_3$)
correspond to the largest threshold
related to a non-vanishing row of $C_2$.
The second channel with wavenumber $\kappa'_2$
and, hence, the second row of the new matrix $C_3$
correspond to the
largest remaining threshold
related to a row of $C_2$ linearly independent of the first row
of $C_2$.
At each next step $i \le R$, a new channel with wavenumber $\kappa'_i$
and row $i$ of $C_3$
corresponds to the largest
remaining threshold related to a row of $C_2$
linearly independent of the previous rows of $C_3$.
The reordering ends when the first $R$ rows of
$C_3$ become linearly independent.
All remaining rows of $C_2$ are transferred to $C_3$
without changes.
As mentioned above we now permute columns in $C_3$ to have
its upper left $R\times R$ block non-singular
thus obtaining matrix $C$. Matrix $D_2$ after
all these permutations is transformed into $D$.

The diagonal wavenumber matrix
is written as
\begin{equation}\label{kps}
\kappa =
\left ( \begin{array}{cc} \kappa' & 0 \\ 0 & \kappa'' \end{array} \right)
\end{equation}
where $\kappa'$ is the $R \times R$ diagonal block
after the reordering
and
$\kappa''$ is the $(N-R) \times (N-R)$ diagonal block containing the remaining
wavenumbers.
Such a structure of $C$, $D$
and wavenumber
matrices will be assumed till the end of
the paper.

Now we can formulate our main theorem.
\begin{Th}
\label{Th:Uinf}
When $r \to \infty$, if rank $C = R$,
the superpotential has the asymptotic form
\begin{eqnarray}\label{Uas}
U(r) \to \mbox{diag}\,(u_{ii}) \\ u_{ii}= + \kappa_i
\qquad
i=1,\ldots,R
\label{UasP}\\
 u_{ii}= - \kappa_i
\qquad
i=R+1,\ldots,N.
\label{UasM}
\end{eqnarray}
\end{Th}

In order to prove the theorem,
we first reduce $C$ and $D$ to  canonical forms,
simplest as far as the superpotential \eref{U} is concerned
but reflecting
on the one hand
 the singular character of
matrix $C$ and
on the other hand
the non-singularity of the whole factorization
solution $\sigma(r)$.
This is performed in the next lemma.

\begin{lemma}
\label{lem:Ccanon}
Matrices $C$ and $D$ can be transformed by right multiplication with a
non-singular
square matrix $T$
into the canonical forms
\begin{equation}\label{CT}
CT = \left ( \begin{array}{cc} I & 0 \\ Q_0 & 0 \end{array} \right)
\qquad
DT = \left ( \begin{array}{cc} X_{0} & -Q_0^T  \\
0 & I \end{array} \right)
\end{equation}
where $I$ denotes
the $R \times R$  unit matrix in $CT$ and
 the $(N-R)\times (N-R)$ unit matrix in $DT$
and $X_0$ is a
symmetric matrix, $X_0^T=X_0$.
Matrix $Q_0$ verifies the following property.
For any $i\le R$ and $j>R$ such that the inequality $\kappa'_{i}<\kappa''_j$
holds one has
\begin{equation}\label{Qij}
q^0_{ji} = 0
\end{equation}
where $q^0_{ji}$ are entries of matrix $Q_0$.
\end{lemma}

\begin{proof}
By construction the $R \times R$ upper left block $M$ of $C$
is invertible.
Since the rank of $C$ is $R$, the last $N-R$ columns of $C$
are linear combinations of the first $R$ ones.
This means that there exists an $R\times (N-R)$ matrix $P$ such that
\begin{equation}
C \left ( \begin{array}{c} P \\ -I \end{array} \right) = 0
\end{equation}
where $I$ is the $(N-R) \times (N-R)$ unit matrix.
Similarly, for the rows,
there exists an $(N-R)\times R$ matrix $Q_0$ such that
\begin{equation}
\left ( \begin{array}{cc} Q_0\, & -I \end{array} \right) C = 0.
\end{equation}
Hence matrix $C$ can be written as
\begin{equation}
C = \left ( \begin{array}{cc} M & MP \\ Q_0M & Q_0MP \end{array} \right)
\end{equation}
or
\begin{equation}
C = \left ( \begin{array}{cc} I & 0 \\ Q_0 & 0 \end{array} \right)
\left ( \begin{array}{cc} M & 0 \\ 0 & I \end{array} \right)
\left ( \begin{array}{cc} I & P \\ 0 & -I \end{array} \right)
\end{equation}
and
\begin{equation}\label{canonC}
CT_1 =
\left(\begin{array}{cc} I & 0 \\ Q_0 & 0 \end{array}
\right)
\qquad
T_1=
\left ( \begin{array}{cc} I & P \\ 0 & -I \end{array} \right)
\left ( \begin{array}{cc} M & 0 \\ 0 & I \end{array} \right)^{-1}.
\end{equation}

 The linear dependence of the rows is not modified by the
multiplication of $C$ by a nonsingular $T_1$ after which row
$i(\le R)$ contains only one non-vanishing element in
column $i$.
Property \eref{Qij} now follows from the fact that
 by construction row $j$ is a linear combination of all
previous rows $i=1,2,\ldots$  of $C$ such that $\kappa_i'>\kappa_j''$.

Instead of matrix $D_2$ the product
\begin{equation}\label{DT1}
DT_1=\left ( \begin{array}{cc} D_{11} &   D_{12} \\
D_{21} &  D_{22} \end{array} \right)
\end{equation}
now appears  in \eref{sigJost}.
According to \eref{symDC} the symmetry property of the
superpotential is translated into the following conditions:
\begin{eqnarray}\label{DQ}
 D_{11}-D_{11}^T=D_{21}^TQ_0-Q_0^T D_{21}\\
D_{12}=-Q_0^TD_{22}
\label{DQ2}.
\end{eqnarray}
Equation \eref{DQ2} together with
 form \eref{canonC} of $CT_1$
implies that if $D_{22}$ in \eref{DT1}
is singular, function
$\sigma(r)$ given in \eref{sigJost} becomes singular for
all $r$, a case we would like to avoid here so that we
necessarily assume $D_{22}$ to be invertible.
Moreover, using \eref{DQ} and \eref{DQ2} we rewrite matrix $DT_1$ as
\begin{equation}\label{DC}
DT_1 = \left ( \begin{array}{cc}X_0 &-Q_0^T \\
 0 & I \end{array} \right)
\left ( \begin{array}{cc} I & 0 \\ D_{21} & D_{22} \end{array} \right)
\end{equation}
where we denoted $X_0=D_{11}+Q_0^TD_{21}$.
Finally, we can define $T=T_1T_2$ where $T_1$
is given by \eref{canonC} and $T_2$ is
the inverted second factor in the right hand
side of \eref{DC}
which keeps unchanged matrix $CT_1$; the condition
$X_0^T=X_0$ follows from \eref{DQ}.
\end{proof}

In the next lemma we establish an important property of
 matrix-valued function $Q(r)$
 which will appear in the asymptotic form of the factorization
 solution.
\begin{lemma}
\label{lem:Z}
For $\kappa'$, $\kappa''$ and $Q_0$ defined as in Lemma \ref{lem:Ccanon},
the $(N-R)\times R $ matrix
\begin{equation}\label{Q}
Q(r) = \rme^{\kappa'' r}Q_0  \rme^{-\kappa' r}
\end{equation}
 tends to zero
when $r$ tends to infinity.
\end{lemma}

\begin{proof}
A matrix element of $Q$ is given by
\begin{equation}
q_{ji}(r) = q^0_{ji}\, \rme^{(\kappa''_j - \kappa'_i) r}
\end{equation}
From Lemma \ref{lem:Ccanon}, either the exponential
tends to zero or coefficient $q^0_{ji}$ vanishes.
\end{proof}


Now we can prove Theorem \ref{Th:Uinf}.

\begin{proof}
According to \eref{eAs},
the factorization solution has the following asymptotic behaviour
\begin{equation}\label{sigJostAsymp}
\sigma(r) \mathop{\to}_{r\rightarrow\infty}\sigma_{\rm as}(r)=
 \kappa^{-1/2}\left[ \rme^{\kappa r}C+ \rme^{-\kappa r} D\right].
\end{equation}
Recalling that
 $C$ and $D$ here have canonical forms \eref{CT} we get
 \begin{equation}\label{SigAs}
\sigma_{\rm as}(r)=\kappa^{-1/2}
\left(
\begin{array}{cc}
I+X& -Q^T \\ Q & I
\end{array}
\right)
\left(
\begin{array}{cc}
\rme^{\kappa'r} & 0 \\ 0 & \rme^{-\kappa''r}
\end{array}
\right)
\end{equation}
where
 $Q$ is given by  \eref{Q}
 and
 matrix
  \begin{equation}\label{P}
X(r)=\rme^{-\kappa'r}X_0\rme^{-\kappa'r}
 \end{equation}
 vanishes at  infinity.
The derivative of  \eref{SigAs} can be written as
\begin{equation}\label{DSig}
\sigma'_{\rm as}(r) =
 \kappa^{1/2}
\left ( \begin{array}{cc} I-X & Q^T   \\
 Q & -I \end{array} \right)
\left ( \begin{array}{cc} \rme^{\kappa' r} & 0 \\
0 & \rme^{-\kappa'' r} \end{array} \right).
\end{equation}
Hence, for $U$ one has the following asymptotic behaviour
\begin{equation}\label{UAS}
\fl U(r) \mathop{\to}_{r\rightarrow\infty} U_{\rm as}(r)  = \kappa^{1/2}
\left ( \begin{array}{cc} I-X & Q^T \\ Q & -I \end{array} \right)
\left ( \begin{array}{cc} I+X & -Q^T \\ Q & I \end{array}\right)^{-1}
\kappa^{1/2}
\end{equation}

From Lemma \ref{lem:Z} and \eref{P} one obtains for $r \to \infty$
\begin{eqnarray}
U_{\rm as}(r)\to
\kappa\left ( \begin{array}{cc} I & 0 \\ 0 & -I \end{array} \right),
\end{eqnarray}
which concludes the proof.
\end{proof}

\begin{cor}
If the thresholds are ordered such that
$\kappa_1>\kappa_2>\ldots>\kappa_N$
and matrix $C$ has rank $R$
the number of arbitrary parameters in the superpotential
is $R(R+1)/2+R(N-R)$.
\end{cor}
\begin{proof}
For the given order of thresholds all $q_{ji}^0$ may be chosen
different of zero
according to Lemma \ref{lem:Ccanon}.
Moreover,
 the parameters enter in the superpotential only
through matrices $X_0$ and $Q_0$.
\end{proof}


\subsection{Jost-matrix transformation}

After the asymptotics of the superpotential is found we are able
to
calculate both the Jost solution and the Jost matrix for the
transformed potential.

\begin{Th}
\label{Th3}
The Jost matrix $\tilde F(k)$ of the transformed potential $\tilde V$ reads,
in terms of the Jost matrix
$F(k)$ and the function $G(k)$ \eref{G}
of the initial potential,
\begin{equation}\label{Ftilde}
\tilde F(k)=\left[U(\infty) - \rmi k\right]^{-1}
\left[F(k)U(0)+G(k)\right].
\end{equation}
\end{Th}

\begin{proof}
According to Theorem  \ref{Th:Uinf}
the Jost solution of the transformed potential reads
\begin{equation}\label{ftilde}
\tilde f(k,r)=A^- f(k,r)\left[U(\infty) - \rmi k\right]^{-1}
\end{equation}
as seen with \eref{Am} and \eref{eAs}.
The theorem then follows from definitions \eref{FJost}
and \eref{G}.
\end{proof}

\section{SUSY partners of $V(r)\equiv0$}

The zero initial potential is  important since in this
case compact analytic expressions are possible both for the
transformed potential and for its Jost function.
The initial Jost solution in this case is simply the exponential
$f(k,r)=\exp(\rmi kr)$ and the initial Jost function is the
identity matrix, $F(k)=I$.
Hence, $G(k)=-\rmi k$.
The regular solution has the form:
$\varphi(k,r)=\sin(kr)k^{-1}$
and the irregular solution is written as
$\eta(k,r)=\cos(kr)$.
The factorization solution respecting Theorem \ref{Th1}
has the form
\begin{eqnarray}
\sigma(r)&=&
\cosh(\kappa r)+\sinh(\kappa r)\kappa^{-1}U(0)   \\ \label{sig0}
&=&
\frac{1}{2}\,\rme^{\kappa r}\left[I+\kappa^{-1}U(0)\right]+
\frac{1}{2}\,\rme^{-\kappa r}\left[I-\kappa^{-1}U(0)\right].
\end{eqnarray}
Another parametrization corresponds to Lemma \ref{lem:Ccanon}
\begin{equation}\label{TF0}
\sigma(r)=
\kappa^{-1/2}\left[ \rme^{\kappa r} C+ \rme^{-\kappa r} D\right]
\end{equation}
with an appropriate choice of matrices $C$ and $D$.

\begin{prop}\label{Prop1}
Let
 $I$ be the $R\times R$ identity matrix,
matrices $X_0$
and
$Q_0$ be chosen according to Lemma \ref{lem:Ccanon},
$Q(r)$ be defined by \eref{Q} and $X(r)$ by \eref{P}.
If the parameters are such that
$
\det Y(r)\ne0\ \forall r\in [0,\infty)
$ where
\begin{equation}
Y(r)=I+X(r)+Q^T(r)Q(r)
\end{equation}
then the potential $\tilde V(r)=-2U'(r)$
with
\begin{equation}\label{Ugeneral}
U=-\kappa+2\kappa^{1/2}
\left(
\begin{array}{cc}
Y^{-1} & Y^{-1}Q^T \\ QY^{-1} & QY^{-1}Q^T
\end{array}
\right)
\kappa^{1/2}
\end{equation}
is a (non-conservative) SUSY partner of $V(r)\equiv0$
and
has the Jost solution
\begin{equation}\label{JostS}
\tilde f(k,r)=
\left[U(r)-\rmi k\right]\rme^{\rmi kr}
\left[U(\infty) - \rmi k\right]^{-1}
\end{equation}
and the Jost function
\begin{equation}\label{JostF}
\tilde F(k)=\left[U(\infty) - \rmi k\right]^{-1}\left[U(0)-\rmi k\right]
\end{equation}
where $U(\infty)$ may be found from (\ref{Uas}-\ref{UasM}).
\end{prop}

\begin{proof}
First we notice that the function
$\sigma(r)=\sigma_{\rm as}(r)$ given in
\eref{sigJostAsymp}
is just function \eref{TF0}
 and, hence, it can be taken as transformation
function to produce a SUSY partner of $V(r)\equiv0$.
For $C$ and $D$ in block forms \eref{CT}
 the superpotential
is given in \eref{UAS}.
From here after some algebra one gets
\eref{Ugeneral}.
Expressions \eref{JostS} and \eref{JostF} for the Jost solution
and the Jost function correspond to \eref{ftilde} and
\eref{Ftilde} for the zero initial potential, respectively.
\end{proof}

\begin{cor}\label{cor1}
For $X_0=0$ the superpotential \eref{Ugeneral} can be written as
\begin{eqnarray}\label{Uu1}
U & = & \kappa-2\kappa^{1/2}Q_r^T\left(Q_rQ_r^T\right)^{-1}Q_r\kappa^{1/2}\\[.5em]
    &  = & -\kappa+2\kappa^{1/2}Q_c\left(Q_c^TQ_c\right)^{-1}Q_c^T\kappa^{1/2}\label{Uu2}
\end{eqnarray}
where
 $Q_r$ and $Q_c$ are  row and  column block
matrices written in terms of $Q$ \eref{Q} as
$Q_r=(Q, -I)$ where $I$ is the $(N-R)\times(N-R)$ identity matrix,
 $Q_c^T=(I, Q^T)$ where $I$ is the $R\times R$ identity matrix.
\end{cor}

\begin{proof}
Using the property
\begin{equation}
Q \left[I+Q^TQ\right]^{-1} = \left[I+QQ^T\,\right]^{-1} Q
\end{equation}
one obtains from \eref{Ugeneral}
\begin{eqnarray}\label{Ur4}
\fl U
& = & -\kappa+
2\kappa^{1/2}\left(
\begin{array}{cc} I & Q^T\\Q & QQ^T\end{array}\right)\!
\left(\begin{array}{cc}I+Q^TQ&0\\0&I+QQ^T\end{array}\right)^{-1}\kappa^{1/2}
\\
\label{Ur3}
\fl & = & \kappa-
2\kappa^{1/2}\left(\begin{array}{cc} Q^TQ & -Q^T\\-Q & I\end{array}\right)\!
\left(\begin{array}{cc}I+Q^TQ&0\\0&I+QQ^T\end{array}\right)^{-1}\kappa^{1/2}.
\end{eqnarray}
Equations \eref{Uu1} and \eref{Uu2}
are nothing but compact forms of
 \eref{Ur3} and
\eref{Ur4} respectively.
\end{proof}

For two particular cases corresponding to rank $C=N-1$
and rank $C=1$ either \eref{Uu1} or \eref{Uu2} takes a
particularly simple form since in one case $Q_r$ is a row and in
the other case $Q_c$ is a column. The explicit expressions are
given in the following corollary:

\begin{cor}\label{cor2}
Let channels be ordered such that
 $\kappa_1>\kappa_2>\ldots>\kappa_{N-1}$.
Let also $Q$ be a row, $Q=(q_{1},\ldots,q_{N-1})$,
$q_i=q^0_i\exp(\kappa_N-\kappa_i)r$ where $q^0_i=0$
 for any $i$ such that $\kappa_N>\kappa_i$ and arbitrary otherwise.
Then
the superpotential has the following block form:
\begin{equation}\label{Urow}
U=\kappa-
\frac{2}{1+QQ^T}\,
\kappa^{1/2}\left(\begin{array}{cc} Q^TQ & -Q^T\\
-Q &1\end{array}\right)\kappa^{1/2}
\end{equation}
where
$Q^TQ$ is an $(N-1)\times(N-1)$ matrix with entries $q_{i}q_{j}$,
$i,j=1,\ldots, N-1$.

Let us now order only the first channel such that
$\kappa_1>\mbox{max}\,(\kappa_2,\ldots,\kappa_{N})$ and
$Q^T$ be a row $Q^T=(q_{2},\ldots,q_{N})$,
$q_i=q^0_i\exp(\kappa_i-\kappa_1)r$.
Then the
superpotential may be written as
\begin{equation}\label{Ucolumn}
U=-\kappa+
\frac{2}{1+Q^TQ}\,
\kappa^{1/2}\left(\begin{array}{cc} 1& Q^T\\
Q & QQ^T\end{array}\right)\kappa^{1/2}
\end{equation}
where
$QQ^T$ is an $(N-1)\times(N-1)$ matrix with entries $q_{i}q_{j}$,
$i,j=2,\ldots,N$.
\end{cor}

\begin{proof}
The statement follows from Lemma \ref{lem:Ccanon} and Corollary
\ref{cor1}. For the first part of the statement rank $C=N-1$
whereas for the second part rank $C=1$.
\end{proof}

Another simplification occurs for rank $C=N$ and a particular
choice of matrix $X_0$.
\begin{cor}
Let ${\cal X}_0$ be a column of $N$ arbitrary real parameters and
${\cal X}=\exp(-\kappa r){\cal X}_0$.
Then the superpotential reads
\begin{equation}\label{Uxi}
U=\kappa-\frac{2}{1+{\cal X}^T{\cal X}}\kappa^{1/2}{\cal X}{\cal X}^T\kappa^{1/2}.
\end{equation}
\end{cor}

\begin{proof}
Choosing
rank $C=N$ (meaning that $Q_0=0$) and
$X_0={\cal X}_0{\cal X}_0^T$,
we have $Y=I+{\cal X}{\cal X}^T$. Using the property
$({\cal X}{\cal X}^T)^2=({\cal X}^T{\cal X}){\cal X}{\cal X}^T$ one gets
$Y^{-1}=I-(1+{\cal X}^T{\cal X})^{-1}{\cal X}{\cal X}^T$.
The statement follows now from \eref{Ugeneral}.
\end{proof}

\section{Examples}

Let us now illustrate the theorems
and Proposition \ref{Prop1}
we have just established
by some exactly-solvable examples,
supersymmetric partners of $V(r)\equiv0$.

\subsection{The $2\times 2$ model with rank $C=2$ (Cox potential)}

Let us start from a two-channel problem.

According to Theorem \ref{Th1}
we choose the transformation function
as given in \eref{sig0}
with the maximal number of arbitrary parameters included in
\begin{equation}\label{w0}
U(0)=\left(
\begin{array}{cc}
\alpha_1 &\beta \\ \beta &\alpha_2
\end{array}
\right).
\end{equation}
Taking \eref{sig0} into account, the condition rank $C=2$ reads
\begin{equation}\label{cond}
(\kappa_1+\alpha_1)(\kappa_2+\alpha_2)-\beta^2\ne0.
\end{equation}
Because of the simple character of the transformation function we
easily find the superpotential according to \eref{U}.
Its off diagonal elements have the form:
$u_{12}=u_{21}=\beta/\det\sigma$.
For the first diagonal element one obtains
\begin{eqnarray}
&\nonumber
u_{11}=\frac{\cosh(\kappa_2 r)}{\det\sigma}
\left[\alpha_1\cosh(\kappa_1 r)+\kappa_1\sinh(\kappa_1 r)\right]
\\
&
+
\frac{\sinh(\kappa_2 r)}{\kappa_2\det\sigma}
\left[\left(\alpha_1\alpha_2-\beta^2\right)\cosh(\kappa_1 r)+
\kappa_1\alpha_2\sinh(\kappa_1 r)\right].
\label{U11Cox}
\end{eqnarray}
Here
\begin{eqnarray}
&
\det\sigma=
\frac{\alpha_2}{\kappa_2}\cosh(\kappa_1 r)\sinh(\kappa_2 r)+
\frac{\alpha_1}{\kappa_1}\cosh(\kappa_2 r)\sinh(\kappa_1 r)
\nonumber
\\
&
+
\cosh(\kappa_1 r)\cosh(\kappa_2 r)+
\frac{\alpha_1\alpha_2-\beta^2}{\kappa_1\kappa_2}\sinh(\kappa_1 r)\sinh(\kappa_2 r)
.
\label{dtCox}
\end{eqnarray}
The element $u_{22}$ is obtained from \eref{U11Cox} by the
replacement  $\kappa_1\leftrightarrow \kappa_2$ and
$\alpha_1\leftrightarrow \alpha_2$.
According to \eref{Vtilde} the transformed potential is simply
twice the derivative of these expressions with the opposite sign.

Since rank $C=2$,
Theorem \ref{Th:Uinf} states that no negative entries
in the asymptotic form of
the superpotential can appear so that
$U(\infty)= \kappa=\mbox{diag}\,(\kappa_1,\kappa_2)$
which can also be checked by a direct calculation.
Applying now Theorem \ref{Th3} we find the Jost matrix for this potential
\begin{equation}\label{JostCox}
\tilde F(k)=\left(
\begin{array}{cc}
\frac{\alpha_1-\rmi k_1}{\kappa_1-\rmi k_1} &
\frac{\beta}{\kappa_1-\rmi k_1}  \\
\frac{\beta}{\kappa_2-\rmi k_2} &
\frac{\alpha_2-\rmi k_2}{\kappa_2-\rmi k_2}
\end{array}
\right)
\end{equation}
We notice that
 up to a change of parameters we obtain one of the
two-channel Bargmann potentials previously found by other means
\cite{cox:64}.

The same superpotential may be rewritten with another parametrization.
According to Proposition \ref{Prop1} for
 rank$\,C=N$ the superpotential reads
  $U=-\kappa+2\kappa^{1/2}(I+X)^{-1}\kappa^{1/2}$
 with $X$ given by \eref{P}
the elements of which  are
 $x_{ij}=\exp(-\kappa_i r-\kappa_j r)x^0_{ij}$, $i,j=1,2$.
 Its more explicit form is
\begin{equation}
\fl U= -\kappa + \frac{2}{(1+x_{11})(1+x_{22})-x_{12}^2}
 \kappa^{1/2}
\left(
\begin{array}{cc}
1+x_{22} & -x_{12} \\-x_{12} & 1+x_{11}
\end{array}
\right)\kappa^{1/2}.
\end{equation}
 After the change of the parameters
\begin{eqnarray}
x^0_{11}&=&\frac{1}{\det[U(0)+\kappa]}
\left[\beta^2-(\alpha_1-\kappa_1)(\alpha_2+\kappa_2)\right]\\
x^0_{22}&=&\frac{1}{\det[U(0)+\kappa]}
\left[\beta^2-(\alpha_1+\kappa_1)(\alpha_2-\kappa_2)\right]\\
x^0_{12}&=&\frac{-2\beta\sqrt{\kappa_1\kappa_2}}{\det[U(0)+\kappa]}
\end{eqnarray}
one recovers the previous result.

\begin{figure}
\begin{center}
\scalebox{0.38}{\includegraphics{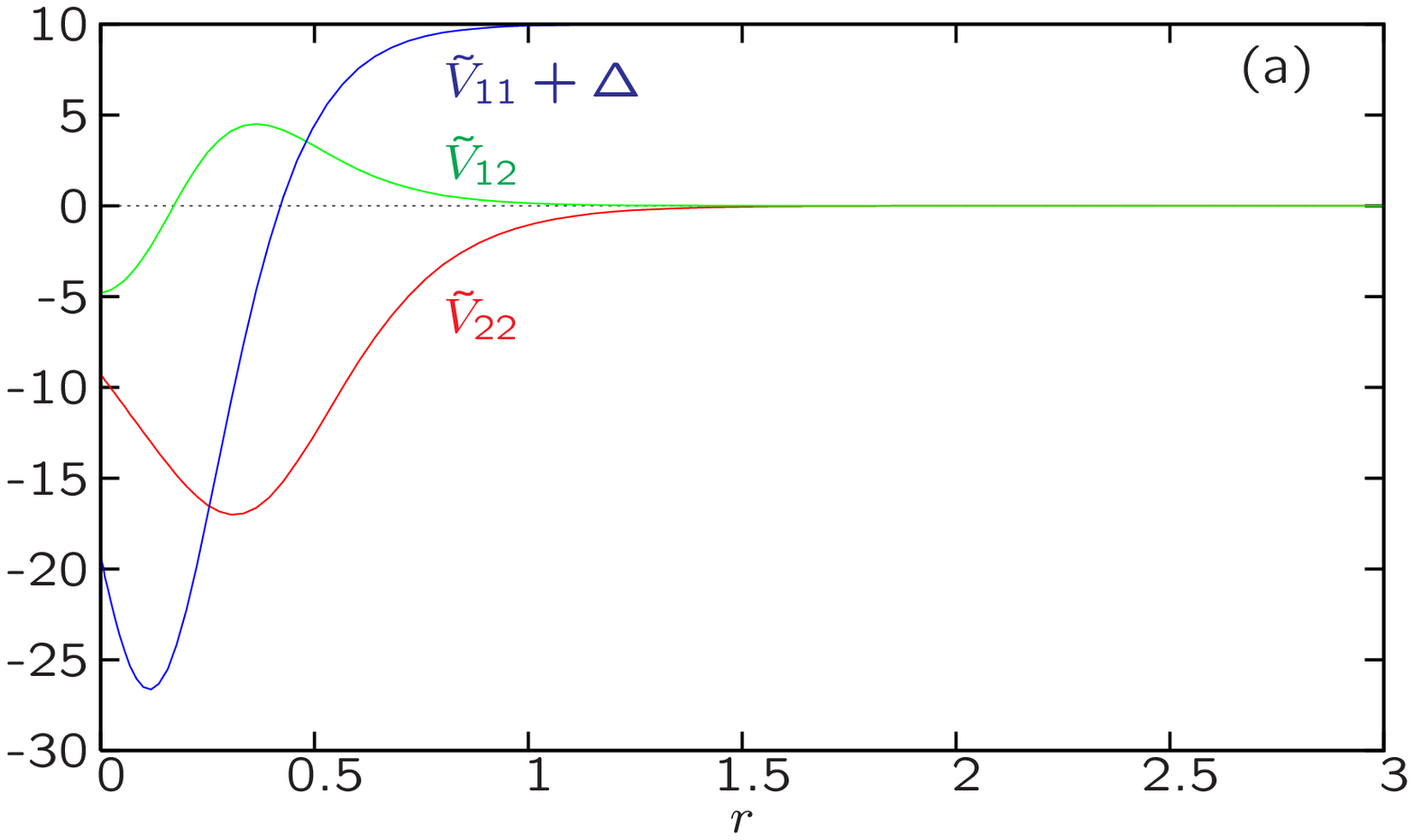}}
\scalebox{0.38}{\includegraphics{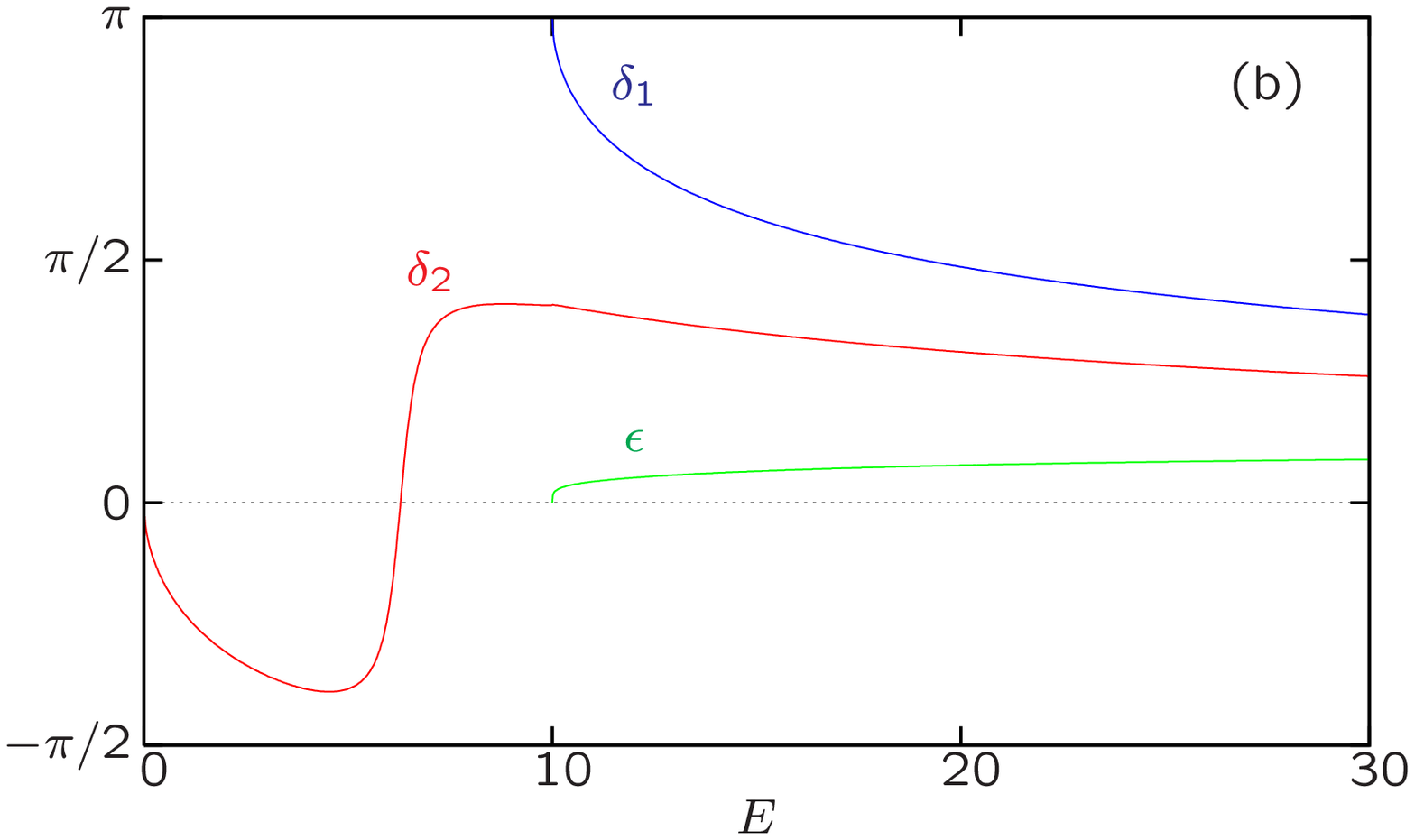}}
 \caption{\label{fig:cox1} $2\times 2$ exactly-solvable potential (a) and scattering matrix (b) for the choice of parameters (\ref{Deltaexample})-(\ref{U0example}) and $\kappa_2=3$ (rank $C=2$).}
\end{center}
\end{figure}

\begin{figure}
\begin{center}
\scalebox{0.38}{\includegraphics{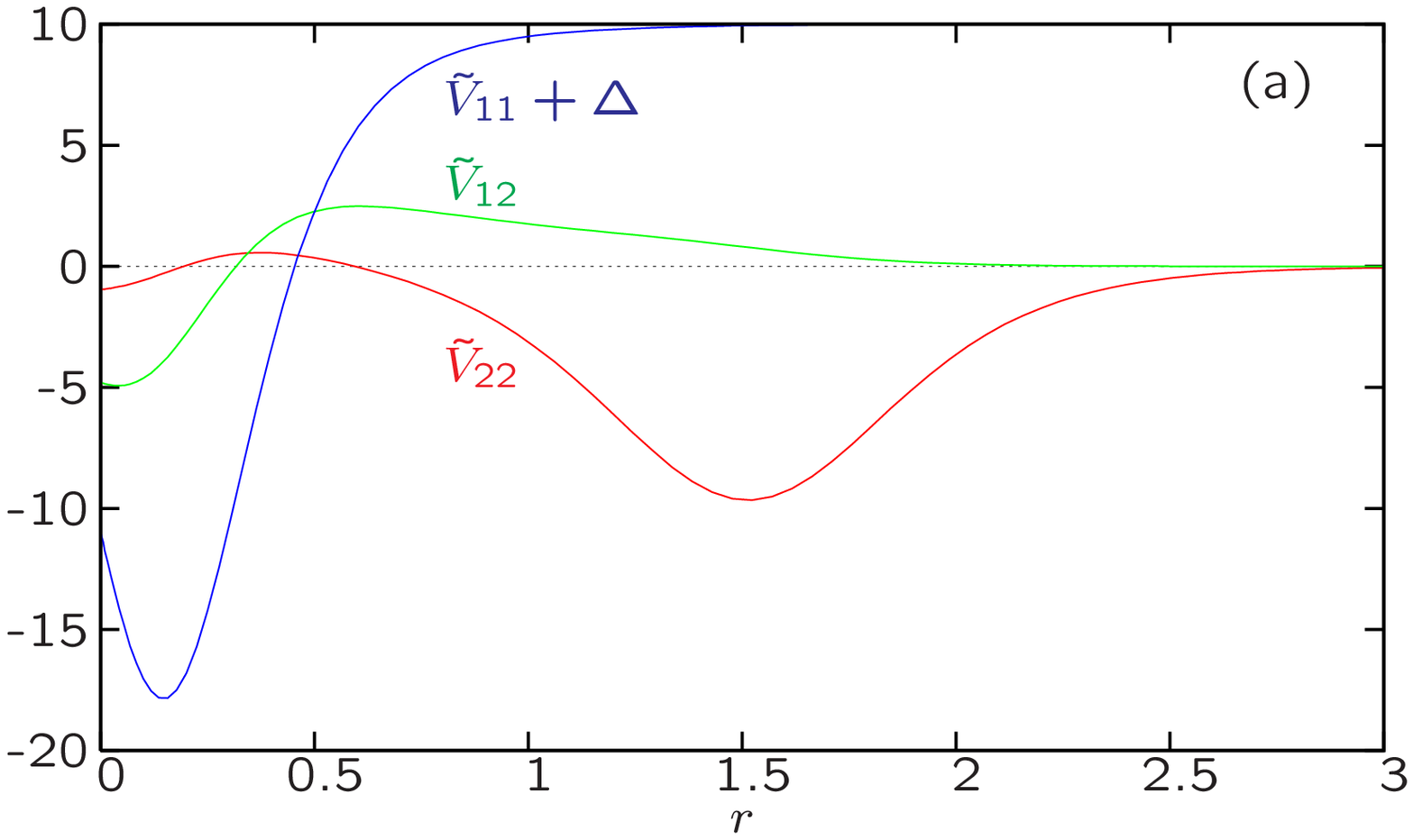}}
\scalebox{0.38}{\includegraphics{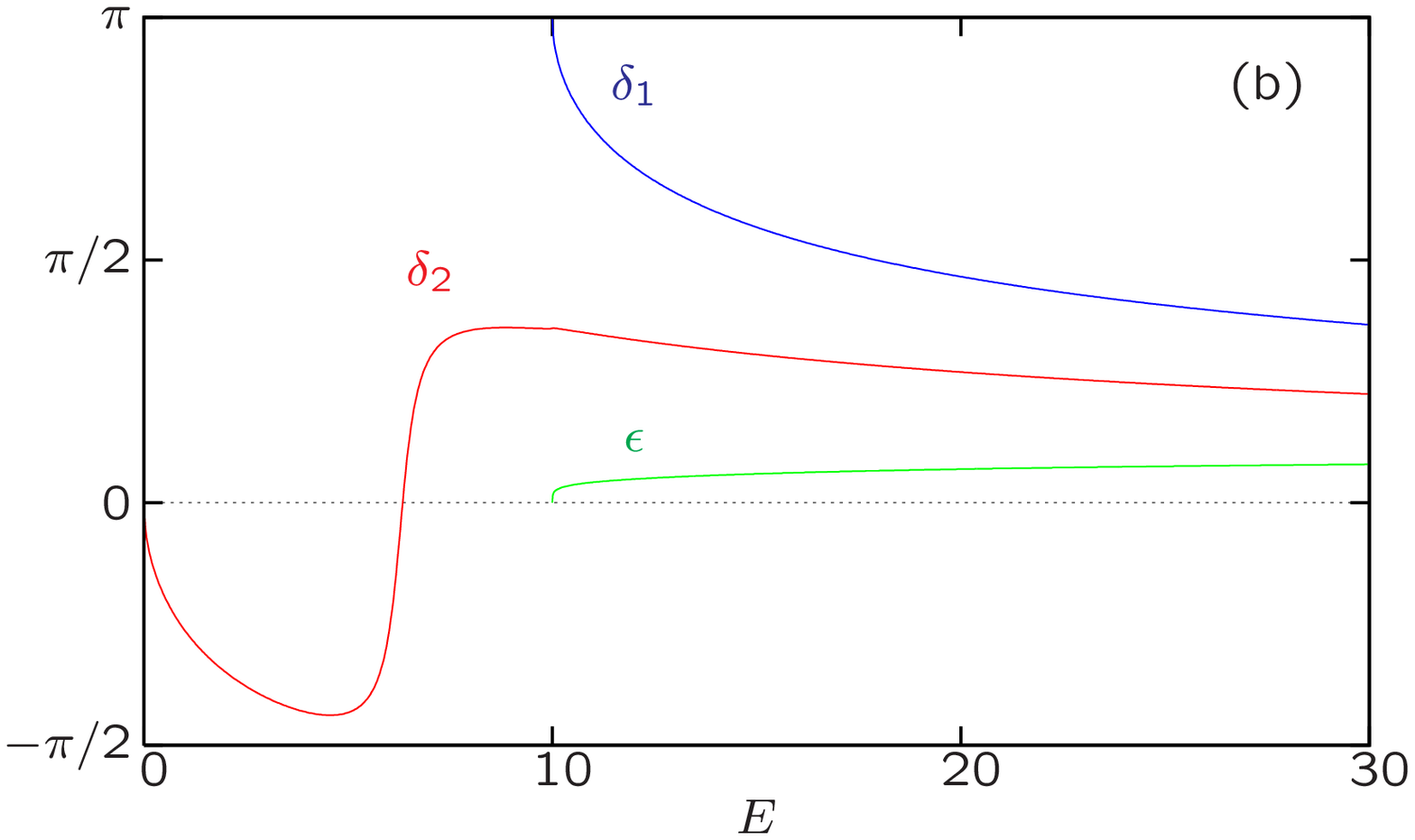}}
 \caption{\label{fig:cox2} Same as figure \ref{fig:cox1} but for $\kappa_2=2.2$ (rank $C=2$).}
\end{center}
\end{figure}

Two explicit examples of this $2 \times 2$ model are given
in figures \ref{fig:cox1} and \ref{fig:cox2}.
Part (a) of each figure shows the potential while part (b) shows the eigenphase shifts $\delta_1$, $\delta_2$ and the mixing parameter $\epsilon$ \cite{newton:82}.
For these examples, we have chosen the thresholds
\begin{eqnarray} \label{Deltaexample}
\Delta_1=\Delta=10, \quad
\Delta_2=0,
\end{eqnarray}
which implies that the factorization energy $\cal E$ and wave number $\{\kappa_1, \kappa_2\}$ are related by
\begin{equation}
\kappa_1^2=\Delta-{\cal E} > \kappa_2^2=-{\cal E}.
\end{equation}
In both cases, the value of the superpotential at the origin, \eref{w0}, is chosen as
\begin{equation} \label{U0example}
U(0)=\left(
\begin{array}{cc}
-2 & 0.6 \\ 0.6 & -2
\end{array}
\right),
\end{equation}
while the factorization wave number is $\kappa_2=3$ and $\kappa_2=2.2$ in figures
\ref{fig:cox1} and \ref{fig:cox2} respectively.
Comparison of figures \ref{fig:cox1} (a) and \ref{fig:cox2} (a) shows that the choice of factorization energy strongly modifies the potential.
In contrast, figures \ref{fig:cox1} (b) and \ref{fig:cox2} (b) show that this choice
does not strongly affect the scattering matrix, which always displays a Feshbach resonance at an energy of about 6.3 and a large negative slope of $\delta_2$ at zero energy and of $\delta_1$ above threshold.
The eigenphase shifts and mixing parameter are only a bit smaller in figure \ref{fig:cox2} (b) than in figure \ref{fig:cox1} (b).

\begin{figure}
\begin{center}
\scalebox{0.38}{\includegraphics{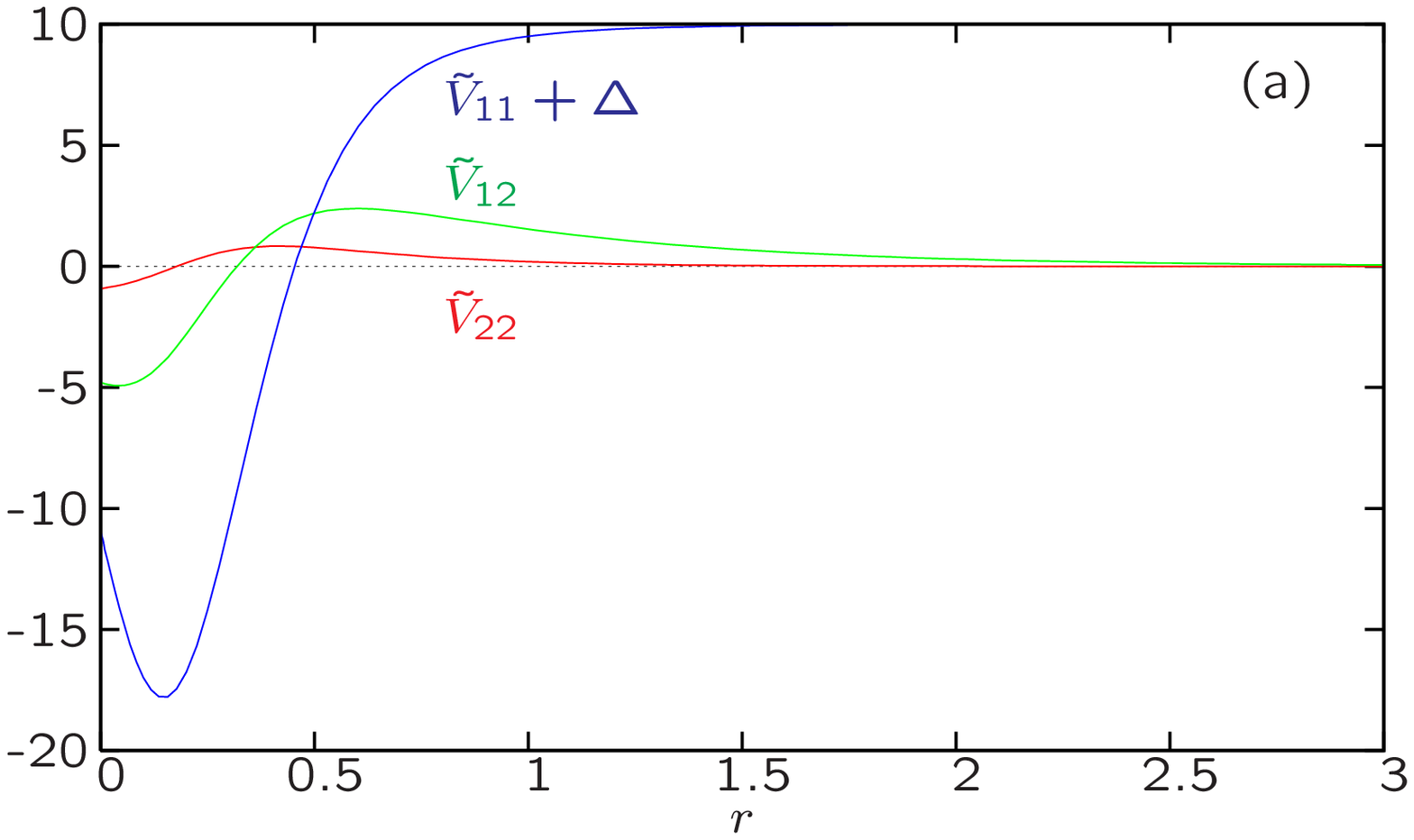}}
\scalebox{0.38}{\includegraphics{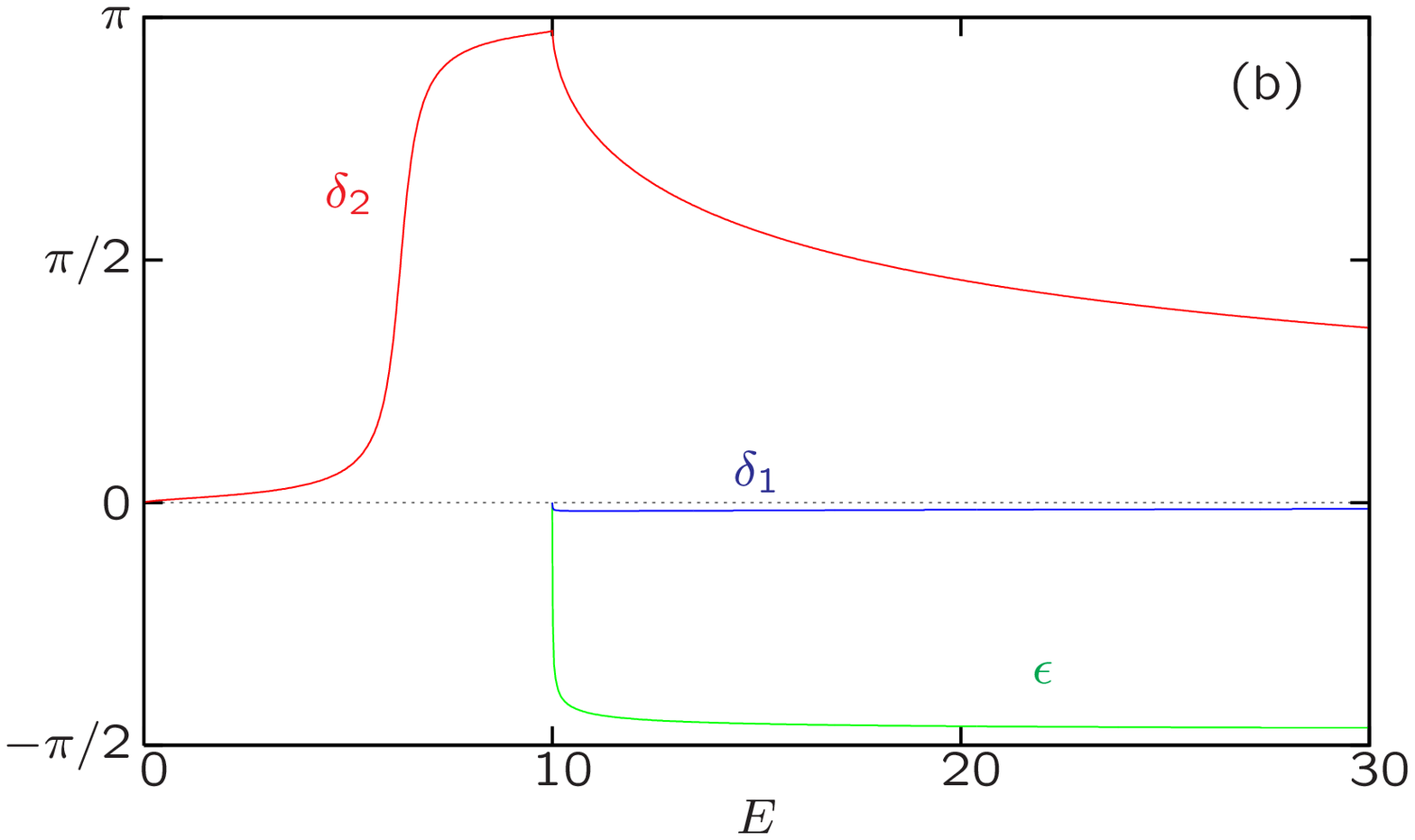}}
 \caption{\label{fig:gcox1} Same as figure \ref{fig:cox1} but for $\kappa_2\approx 2.194\,675\dots$ (rank $C=1$).}
\end{center}
\end{figure}

\subsection{The $2\times 2$ model with rank $C=1$}

In the previous $N=2$ example, matrix $C$ has rank 2.
Let us now consider the case, not allowed
in \cite{cox:64}, where rank $C$ is 1.
Our supersymmetric formalism, on the contrary, is also valid in this case.
Both matrices $Q_0$ and $X_0$
defined in Lemma \ref{lem:Ccanon}
are  numbers, $Q_0\equiv q_0$, $X_0\equiv x_0$,
so that $Q(r)\equiv q(r)=q_0\exp(\kappa_2 r-\kappa_1 r)$ and
$X(r)\equiv x(r)=x_0\exp(-2\kappa_1r)$.
Choosing $\kappa_1>\kappa_2$ and $x_0>-1-q_0^2$ we get
from \eref{Ugeneral}
\begin{equation}\label{Uex1}
U=\left(
\begin{array}{cc} a_1 & b \\ b & a_2 \end{array}
\right)
\end{equation}
with
\begin{equation}
\fl a_1=\frac{1-x-q^2}{1+x+q^2}\,\kappa_1 \qquad
a_2=-\frac{1+x-q^2}{1+x+q^2}\,\kappa_2 \qquad
b=\frac{2q\sqrt{\kappa_1\kappa_2}}{1+x+q^2}.
\end{equation}
It is not difficult to check
the condition rank $C=1$ (cf. \eref{cond})
\begin{equation}\label{condaa}
(\kappa_1+\alpha_1)(\kappa_2+\alpha_2)-\beta^2=0
\end{equation}
 where $\alpha_{1,2}=a_{1,2}(0)$ and $\beta=b(0)$.
 We notice that our approach permits us to calculate
the superpotential by the same formulas
\eref{U11Cox} and \eref{dtCox} as in the previous section
 but now
the parameters are not independent anymore,
they should satisfy condition \eref{condaa}.
With the identification $x_0=0$ and
$4\kappa_1\kappa_2/\beta^2=(q_0+1/q_0)^2$
this leads to an exactly-solvable model used in \cite{sparenberg:06}
to construct an analytical model for the Feshbach-resonance phenomenon.

Theorem \ref{Th1} implies that
$U(\infty)=\mbox{diag}\,(\kappa_1,-\kappa_2)$
which is clearly seen from \eref{Uex1}.
The value $U(0)$ is given by the same formula \eref{Uex1} with
the replacement $x\to x_0$ and $q\to q_0$. Therefore applying
Proposition \ref{Prop1} one gets the Jost matrix
\begin{equation}\label{JostEx1}
\tilde F(k)=
\left(
\begin{array}{cc}
\frac{\alpha_1-\rmi k_1}{\kappa_1-\rmi k_1} &
\frac{\beta}{\kappa_1-\rmi k_1}\\[.5em]
-\frac{\beta}{\kappa_2+\rmi k_2} &
-\frac{\alpha_2-\rmi k_2}{\kappa_2+\rmi k_2}
\end{array}
\right)
\end{equation}
which differs from \eref{JostCox} because of the different asymptotic
form of the superpotential at infinity.

Let us now construct an explicit example that illustrates the strong impact of
$U(\infty)$ on the scattering matrix.
In figure \ref{fig:gcox1}, a potential is constructed with the same parameters as in figures \ref{fig:cox1} and \ref{fig:cox2}, except for the factorization wave number which is now chosen in order to satisfy condition (\ref{condaa}),
$\kappa_2 \approx 2.194\,675\dots$
When $\kappa_2$ is smaller than this limit,
the potential becomes singular, a case we want to avoid here.
Figures \ref{fig:cox1} (a), \ref{fig:cox2} (a) and \ref{fig:gcox1} (a) show that
when $\kappa_2$ approaches this limit value from above, the well in potential $\tilde{V}_{22}$ goes to infinity and finally disappears when $\kappa_2$ actually reaches the limit value.
As long as $\kappa_2>2.194\,675\dots$,
this well movement in $\tilde{V}_{22}$ has practically no impact on the scattering matrix: the potentials are nearly phase equivalent with each other.
In contrast, when the well disappears,
a strong change of behaviour is observed, as seen in figure \ref{fig:gcox1} (b):
while the Feshbach resonance still keeps the same energy and width,
the slope of $\delta_2$ at zero energy becomes small and positive.
Above threshold, $\delta_2$ has now a value very close to $\delta_1$ of figures \ref{fig:cox1} (b) and \ref{fig:cox2} (b), with a large negative slope.
Though $\delta_2$ is continuous at threshold, it now has a strong cusp effect.
The mixing parameter, which was close to zero, now gets close to $-\pi/2$,
while $\delta_1$ gets very small.
The rank of $C$, and hence $U(\infty)$, thus have a strong qualitative effect on the scattering matrix, which displays very different behaviours in both cases;
for practical applications, e.g.\ for fitting actual scattering data with such potentials,
both behaviours might be of interest.

Let us finally remark that the behaviour
 $U(\infty)=\mbox{diag}\,(\kappa_1,-\kappa_2)$
agrees with the result obtained in \cite{amado:88a} for $N=2$
 in the context of a bound-state removal by supersymmetric quantum mechanics.
There, the factorization matrix solution is made of one increasing
and one decreasing vector solution at infinity, which corresponds to rank $C$ = 1.
In  \cite{amado:88a}, this result is even generalized to $N$ channels
and a diagonal matrix for $U(\infty)$ is found,
with all positive elements but one,
corresponding to the channel with the lowest threshold.
However, the general situation is more complicated,
as described by our Theorem \ref{Th:Uinf}:
there may be more than one negative element
in the asymptotic form of
the superpotential
and the lowest channel may correspond to a positive asymptotic
entry of the superpotential.
This possibility is illustrated by our last example.

\subsection{The $3\times3$ model with rank $C=2$ \label{ex3}}

Here we choose
$\kappa_1>\kappa_3>\kappa_2$,
rank $C=2$ so that matrix $Q_0$ is the
row $Q_0=(q_0,0)$.
Such a choice of $Q_0$ reflects the fact that the third row of $C$
is proportional to its first row
(see \eref{Qij}).
The adopted order of
thresholds corresponds to $\kappa'=\mbox{diag}\,(\kappa_1,\kappa_2)$ and
$\kappa''=\kappa_3$.
For simplicity
only (equal) off-diagonal entries of
$2\times2$ matrix $X_0$ from Proposition \ref{Prop1}
are chosen different of zero which we denote  $x_0$.
Then applying \eref{Ugeneral} we obtain
\begin{equation}
U=-\kappa+\frac{2}{\det\sigma}\,\kappa^{1/2}
\left(
\begin{array}{ccc}
1 & -x & q\\
-x &1+q^2 & -xq\\
q & -xq & q^2
\end{array}
\right)\kappa^{1/2}
\end{equation}
where $\det{\sigma}=1+q^2-x^2$,
$x=x_0\exp(-\kappa_1r-\kappa_2r)$, $q=q_0\exp(\kappa_3r-\kappa_1r)$.
The condition
$\det\sigma\ne0$
$\forall r\in[0,\infty)$ is satisfied if
$q_0^2>x_0^2-1$.
It is clearly seen here that at $r\to\infty$ both $q$ and $x$
vanish and the asymptotic form of $U$,
$U_{\rm as}=\mbox{diag}\,(\kappa_1,\kappa_2,-\kappa_3)$,
agrees with Theorem
\ref{Th:Uinf}.


\section{Conclusion and perspectives}

In conclusion, supersymmetric quantum mechanics is a promising tool
to pragmatically solve the inelastic coupled-channel inverse problem,
thanks to its ability to construct sophisticated exactly-solvable models.
The present work is only a starting point to the general study
of supersymmetric transformations in the multichannel case:
by focusing on the asymptotic behaviour of the superpotential,
it reveals the richness of possible behaviours,
as compared with the single-channel case.

We have made a rigorous study of this asymptotic behaviour for a
general supersymmetric transformation.
Our theorem generalizes a result found in the literature
\cite{amado:88a,amado:88b,cannata:93}:
there may be several negative elements in the asymptotic form of the
superpotential and the lowest threshold may
produce either a positive
(as shown explicitly by our example \ref{ex3})
or a negative entry.

As a byproduct of our proof, we have been able to construct exactly-solvable
potentials that are supersymmetric partners of the zero potential.
Essential simplifications
occur when each element of the
factorization matrix solution is either an increasing or a decreasing
exponential (not a linear combination of both types).
Explicit examples have been given for two and three channels.

Future work should focus on both the behaviour of the superpotential
at the origin and at infinity in the general case:
arbitrary number of channels, with equal and/or different thresholds,
and arbitrary rank for the matrices multiplying the regular and singular
solutions in the factorization matrix solution.
On the other hand, iterations of coupled-channel
supersymmetric transformations should be studied,
which should eventually lead to a satisfactory solution
of the coupled-channel inverse problem, with and without threshold
difference.

\ack
We acknowledge useful discussions with Fran\c{c}ois Foucart and Andrey Pupasov,
at the beginning and at the end of this work respectively.
This text presents research results of the Belgian program P5/07
on interuniversity attraction poles of the
Belgian Federal Science Policy Office.
BFS is partially supported by grants RFBR-06-02-16719 and
SS-5103.2006.2 and thanks the National Fund for Scientific Research, Belgium,
for support during his stay in Brussels.


\end{document}